\documentclass{webofc}

\usepackage[varg]{txfonts}   
\usepackage{hyperref}
\usepackage{url}

\usepackage{import} 
\usepackage{siunitx} 
\hypersetup{colorlinks=true,citecolor=blue,urlcolor=blue,linkcolor=blue}
\begin{document}
\title{Neutrino mass experiments: current and future}
%
%

\author{\firstname{Larisa A.} \lastname{Thorne}\inst{1}\fnsep\thanks{\email{lthorne@uni-mainz.de}}
}

\institute{Johannes Gutenberg University Mainz
}

\abstract{Nearly 70 years since the neutrino was discovered, and 25 years since discovery of neutrino oscillations established its non-zero mass, the absolute neutrino-mass scale remains unknown. 
Due to its unique characteristics, determining this neutrino property requires new measurement techniques to be developed. Currently, there are four measurement approaches: using cosmological models, inference from time-of-arrival from supernovae, through observation of neutrinoless double beta decay, and the kinematics of weak decay processes.

I will review the theoretical basis underlying neutrino mass measurement and present key experiments in this field. I will highlight the current best upper limits, how neutrino mass experiments are complementary to other neutrino property searches, and summarize the challenges that lie ahead of the neutrino mass community.
}
\maketitle
\section{Introduction}
\label{intro}

In the many decades since the discovery of the neutrino, fundamental questions relating to its nature remain unanswered. There has been significant progress in that time, thanks to international investment in developing new methods and detection techniques. 

One of these fundamental questions about neutrino properties (described in Section 2) is the question of their absolute mass scale. There are a variety of techniques to measuring the absolute mass scale. Section 3 describes these in detail, and reports on the most recent activity in the field. Summaries of the complete histories are given in \cite{nuMassRev,nucciotti2016,cosmo,Agostini_2023}.

\section{Neutrinos and their properties}
\label{sec:prop}

The neutrino sector has several interesting features, many stemming from neutrino oscillations. 

Neutrino oscillations govern the identity of the particle. Neutrinos are born in weak decay processes, with a flavor eigenstate matching that of its leptonic decay partner: either an electron, muon, or tau particle. The flavored neutrino will then propagate through spacetime in the mass eigenstate, which are referred to as \{1, 2, 3\}. These mass eigenstates are an incoherent sum of the flavor eigenstates \{e, $\mu$, $\tau$\}, weighted by the PMNS mixing matrix. By construction, the mass eigenstate 1 is the state with the largest admixture of the electron-flavor eigenstate. Finally, when the neutrino is observed sometime later, it is detected in its flavor eigenstate. Due to this identity oscillation during propagation, the same neutrino can be detected in a different flavor than it was born with.

From neutrino oscillations, the probability that a neutrino born in flavor eigenstate \textit{i} is detected in flavor eigenstate \textit{j} is proportional to difference of mass eigenvalues squared, $m_i^2-m_j^2$. Interesting questions arise here, such as ordering of these mass eigenvalues relative to one another, their absolute mass scale, and whether the neutrino is its own anti-particle. Experiments which measure only the oscillation properties (``oscillation experiments") can only offer incomplete answers to these mass-related questions, necessitating dedicated neutrino mass measurement experiments.
\section{Neutrino mass measurements with current experiments}
\label{sec:currExp}

There are four approaches to measuring the absolute mass scale of the neutrino.

\subsection{Cosmology}

This approach relies on fitting various models to cosmological data, including Cosmic Microwave Background (CMB), Baryon Acoustic Oscillation (BAO), and Big Bang Nucleosynthesis (BBN) datasets. The output value is the sum of neutrino mass eigenvalues, $M_\nu = \sum_{i=1}^3 m_i$. The advantage to this approach is the varied, complementary datasets. However, the most significant challenge is the strong model dependence. Various resultant neutrino mass limits are tabulated in Table 2 of a recent summary of the field~\cite{cosmo}, ranging from $M_\nu <$ \SI{0.0866}{\electronvolt} for a standard recipe with no additional extensions in the degenerate hierarchy, to $M_\nu <$ \SI{0.265}{\electronvolt} with an $w_a+w_0$ extension in the inverted hierarchy. Both these upper limits are at 95\% confidence.

A few months after this summary article was published, the DESI collaboration published an analysis of its first year of BAO observations ("DESI Data Release 1")~\cite{desiRelease1}. This includes a new $M_\nu < 0.072~(0.113)$ \SI{}{\electronvolt} for normal (inverted) hierarchy, at 95\% confidence. A separate group~\cite{yeung2024} which analyzed the same dataset claimed that including a non-zero free value of $M_\nu$ and a degeneracy parameter $\xi_3$ in their models could resolve long-standing Hubble constant tensions.

\subsection{Supernova time-of-flight}

The dispersion in time-of-arrival of neutrinos from a single source is dependent on neutrino mass. The best way to apply this method is to observe supernovae, which are an excellent source of neutrinos. The observable here is the electron-weighted neutrino mass squared ($m_{\nu_{e}}^2 = \sum_{i=1}^3|U_{e,i}|^2 m_i^2$), as the measured neutrinos are electron-flavored.

The most successfully studied supernova is also the most recent one: Supernova 1987A. Four experiments (Kamioka, Baksan, Mt. Blanc, and Homestake) collectively measured the timestamps and energies of around 25 neutrinos from this event~\cite{schramm}, and a later estimate on the neutrino mass limit was $m_{v_{e}} <$ \SI{5.7}{\electronvolt} at 95\% confidence~\cite{supernova}. 

There are many advantages to this technique: the large number of detectors, and the additional information that can be extracted from the data, including mass hierarchy information via MSW effect on the 1-3 mixing, and information on stellar structure and the equation of state. The challenges are the low statistics (i.e., having the luck to be ready for the next supernova event) and the fact that the optimal signal is not the main detection channel for most detectors (i.e., lower detection efficiency).

\subsection{Search for neutrino-less double beta decay}

There are many processes which produce (or consume) neutrinos which can be studied. One of these is neutrino-less double beta decay, a process which involves two neutrinos annihilating one other. If this process is observed, a decay rate can be calculated and the neutrino mass can be estimated using Fermi's Golden Rule. The key is that, in order for the process to occur, neutrinos must be their own anti-particle (type "Majorana"), and so the interpretation of the neutrino mass is different because it includes additional parameters: $m_{\beta \beta} = \big| \sum_{i=3}^3 |U_{e,i}|^2 m_{i} e^{i \alpha_i} \big|$. 

One of the major advantages of this mass measurement method is that there are a multitude of candidate isotopes, detectors, and signal readout techniques; they are described extensively in \cite{Agostini_2023}, and will be summarized below. There are around 40 candidate isotopes for neutrino-less double beta decay, the top three most common being $^{76}$Ge, $^{136}$Xe, and $^{100}$Mo. There are four main detection channels, which determines event reconstruction technique: ionization, scintillation light, phonon, and Cerenkov light. There are no current experiments which rely on Cerenkov light, but it has been proposed for future experiments or to improve understanding of individual event topology when used in tandem with another method.

Experiments using the ionization detection channel involve measurement of the movement of charge. These are divided into two main groups of experiments: "HPGe" (high purity germanium) which measures charge currents in a semiconductor device constructed with up to 92\% $^{76}$Ge crystal (experiments: GERDA, MAJORANA, LEGEND); and "Xenon TPCs" (time projection chambers) which measure charge drift times through liquid or gaseous chambers to reconstruct events from $^{136}$Xe decays (experiments: EXO-200, nEXO, NEXT, PANDA-X, LZ, and DARWIN).

Experiments which are scintillation-based measure light information, and are particularly suited for reconstructing a complete picture of the event, including timing, position, and energy information carried by photons. They can also be used concurrently with other detection channels; for example, to provide timing information in a TPC experiment. This isn't limited to a particular isotope: the KAMLAND-Zen experiment uses an enriched $^{136}$Xe-filled balloon, whereas the SNO+ experiment will have a $^{130}$Te-filled acrylic sphere.

Phonon-based experiments measure the energy of phonon absorption in a given material, so it is a calorimetric method. There are two main categories of this kind of experiment: tracking and cryogenic. The tracking method is sensitive to a complete picture: it can track phonon deposition energy and the angular distribution (experiments: NEMO-3 with $^{100}$Mo, Super-NEMO with $^{82}$Se). The cryogenic method employs bolometers operated around  \SI{10}{\milli \kelvin} temperatures, and their small size is advantageous due to the granularity with which they can reconstruct events, if deployed as part of a large array. For this method, the isotope needs to be trapped within a crystal so it can be coupled to a thermal sensor. Experiments and isotope preparations are diverse: CUORE used TeO$_2$; CUPID-0 used ZnSe; and CUPID-Mo, CROSS, AMoRE-II, and CUPID used Li$_2$MoO$_4$.

The current best neutrino mass upper limit from neutrino-less double beta decay is $m_{\beta \beta} <$ 36-156 \SI{}{\milli \electronvolt} ~\cite{Agostini_2023,kz_mbb}. Besides the many approaches and individual experiments, the advantage to this method for neutrino mass measurements is that it addresses the question of whether the neutrino is its own antiparticle. The challenges are the significant investment in background identification and suppression methods, the large systematic uncertainty in the nuclear matrix element calculation (this is why the upper limit is quoted as a range), and the unknown phase parameters $\alpha_i$.

\subsection{Kinematic methods (single beta decay)}

Single beta decay, in which one neutron of the parent particle converts into a proton, plus an outgoing electron and an outgoing electron-flavored neutrino, is a direct method for measuring neutrino mass. By measuring the number and energy of the outgoing electrons (``beta" particle), the beta decay spectrum can be obtained. This spectrum can be modelled analytically using Fermi's Golden Rule, with at minimum 4 free fit parameters: the maximum beta energy (``endpoint") $E_0$, the effective electron-weighted neutrino mass squared $m^2_{\beta}$, a normalization $A$ which scales with source activity, and a background term $B$. By fitting the measured beta spectrum with this model, the neutrino mass $m^2_{\beta}$ can be extracted, where $m_{\beta}^2 = \sum_{i=1}^3|U_{e,i}|^2 m_i^2$.

There are a few experiments which use this approach, but with a wide variety of techniques. Several experiments use a tritium ($^3$H) beta decay isotope, which has an endpoint energy of \SI{18.6}{\kilo \electronvolt} and a half-life of 12.3 years. Past experiments, such as the Los Alamos\cite{losAlamos}, Mainz\cite{mainz2005}, and Troitsk\cite{troitsk2011} experiments, provided an important foundation for the experiments of today. The KATRIN\cite{hardwarepaper} experiment uses an ultra-luminous molecular tritium beta decay source, and measures the integrated beta decay spectrum using a MAC-E filter and segmented silicon p-i-n diode detector. It has a design sensitivity of \SI{0.3}{\electronvolt}, and is currently gathering data until 2026. There have been several intermediate data releases, including the latest one~\cite{katrin2022} which set a new upper limit of $m_{\beta}<$ \SI{0.8}{\electronvolt} at 90\% confidence\footnote{Update: a few weeks after this conference, the KATRIN collaboration released a new result with a larger dataset, which shrank the upper limit to $m_{\beta}<$ \SI{0.45}{\electronvolt} at 90\% confidence~\cite{katrin2024}}. 

There are two other tritium experiments in the pipeline. The Project 8\cite{oldp8} experiment traps gaseous atomic tritium in a magnetic trap, and measures the differential beta spectrum by extracting kinetic energy information from the cyclotron radiation emitted from the trapped electrons ("CRES" technique). This next-generation experiment has successfully produced and analyzed a molecular tritium demonstrator experiment, with an extracted upper limit on the neutrino mass of $m_{\beta} <$ 152 (155) \SI{}{\electronvolt} at 90\% confidence for a frequentist (Bayesian) analysis\cite{p8PRL}. In order to achieve the ultimate design sensitivity of \SI{40}{\milli \electronvolt}, the limiting systematic contributions as well as the statistics must be addressed. In terms of systematics, the leading contribution is from the final states of molecular tritium: here the probabilities of transferring energy into particular rotational, vibrational, and electronic excitation states of the molecule results in a smearing of the beta decay spectrum. Using atomic tritium circumvents the issue, but creating and maintaining a cold tritium atoms (tritons) is challenging, and development is ongoing. To address the limited available statistics, the magnetic trap volume must increase while simultaneously maintaining a strict magnetic shaping and homogeneity requirement; development is ongoing in this area as well. 

The final tritium experiment is PTOLEMY\cite{ptolemy}, which envisions an amalgamation of the best features of all current tritium neutrino mass experiments. When constructed, PTOLEMY will have a \SI{100}{\gram} tritiated graphene source, a electromagnetic filtering spectrometer with radio frequency tracking capabilities, and a transition-edge sensor (TES) microcalorimeter to detect. These large quantities are required because PTOLEMY aims to measure relic neutrinos; these relic neutrinos will appear as a small peak centered at a distance $2m_{\beta}$ above the beta decay endpoint.

Other beta decay isotopes besides tritium are available. There is much interest in inverse beta decay (or electron-capture) on $^{163}$Ho, which has an endpoint energy of \SI{2.8}{\kilo \electronvolt} and a half-life of 4570 years. The main active experiments are ECHo\cite{echo2019} and Holmes\cite{BORGHESI2023}, previously also NuMECS. These experiments primarily rely on calorimetric measurements of deposited electron energy (using arrays of metallic magnetic calorimeters (MMC) in the ECHo experiment, and superconducting TES units in the Holmes experiment). Of the holmium-based experiments, the best upper limit is $m_{\beta} <$ \SI{150}{\electronvolt} at 95\% confidence, by the ECHo collaboration. 

There are two additional isotopes of interest: $^{187}$Re and $^{241}$Pu. There is no current active work in rhenium, as the material proved challenging to work with. The plutonium research branch is in its infancy, but appears to be more well-suited for sterile neutrino searches.

In summary, of the various single beta decay experiments, the best upper limit is $m_{\beta} <$ \SI{0.8}{\electronvolt} at 90\% confidence (tritium isotope with the KATRIN experiment). The advantages of using single beta decay as a means to extracting neutrino mass is the myriad of experiments with different techniques and different isotopes, as well as cross-checks with other experiments on intermediate results, like isotope Q-values with Penning trap experiments. Some of the challenges which must be addressed by future experiments are increasing statistics by scaling up, and addressing systematic effects, like molecular final states and backgrounds.

\section{The future}
\label{sec:future}

In order to finally make a direct measurement of the neutrino mass, rather than setting new upper limits, the neutrino mass community must make a few key investments in statistics, controlling systematics, combining analyses, and gaining new information from complementary searches. 

To improve statistics, the experiments must be scaled up. This can be accomplished either through building much larger experiments, developing modular units which can be mass-produced, increasing source activity, or increasing measurement time duration. Systematics can be understood through dedicated measurement campaigns or by collaboration with theorists (in the case of nuclear matrix calculation for neutrino-less double beta decay, or molecular final state calculation for single tritium beta decay). In addition, much can be learned if two similar experiments perform a combined analysis, as is common in other fields. In a similar vein, complementary searches can inform or constrain neutrino mass inputs; examples include the Q-values from Penning trap experiments, which directly impact the single beta decay endpoint (a parameter which is strongly correlated with the neutrino mass squared), or sterile neutrino searches.


\begin{thebibliography}{20}

\bibitem{nuMassRev}
J.A. Formaggio, A.L.C. de~Gouv{\^{e}}a, R.H. Robertson, Direct measurements of neutrino mass, Physics Reports \textbf{914}, 1 (2021). \doiwoc{10.1016/j.physrep.2021.02.002}

\bibitem{nucciotti2016}
A.~Nucciotti, {The Use of Low Temperature Detectors for Direct Measurements of the Mass of the Electron Neutrino}, Advances in High Energy Physics \textbf{vol. 2016}, pg. index 9153024 (2016), \texttt{https://onlinelibrary.wiley.com/doi/pdf/10.1155/2016/9153024}. \doiwoc{https://doi.org/10.1155/2016/9153024}

\bibitem{cosmo}
E.D. Valentino, S.~Gariazzo, O.~Mena, Neutrinos in cosmology (2024), \texttt{2404.19322}, \urlstyle{tt}\url{https://arxiv.org/abs/2404.19322}

\bibitem{Agostini_2023}
M.~Agostini, G.~Benato, J.A. Detwiler, J.~Menéndez, F.~Vissani, Toward the discovery of matter creation with neutrinoless beta decay, Reviews of Modern Physics \textbf{95} (2023). \doiwoc{10.1103/revmodphys.95.025002}

\bibitem{desiRelease1}
A.G. Adame, J.~Aguilar, S.~Ahlen, S.~Alam, D.M. Alexander, M.~Alvarez, O.~Alves, A.~Anand, U.~Andrade, E.~Armengaud et~al., {DESI 2024 VI: Cosmological Constraints from the Measurements of Baryon Acoustic Oscillations} (2024), \texttt{2404.03002}, \urlstyle{tt}\url{https://arxiv.org/abs/2404.03002}

\bibitem{yeung2024}
S.~Yeung, W.~Zhang, M.C. Chu, {Resolving the $H_0$ and $S_8$ tensions with neutrino mass and chemical potential} (2024), \texttt{2403.11499}, \urlstyle{tt}\url{https://arxiv.org/abs/2403.11499}

\bibitem{schramm}
D.N. Schramm, {Neutrinos from supernova 1987A}, Comments on Nuclear and Particle Physics \textbf{17}, 239 (1987).

\bibitem{supernova}
T.J. Loredo, D.Q. Lamb, {Bayesian analysis of neutrinos observed from supernova SN 1987A}, Physical Review D \textbf{65} (2002). \doiwoc{10.1103/physrevd.65.063002}

\bibitem{kz_mbb}
S.~Abe, S.~Asami, M.~Eizuka, S.~Futagi, A.~Gando, Y.~Gando, T.~Gima, A.~Goto, T.~Hachiya, K.~Hata et~al. (KamLAND-Zen Collaboration), {Search for the Majorana Nature of Neutrinos in the Inverted Mass Ordering Region with KamLAND-Zen}, Phys. Rev. Lett. \textbf{130}, 051801 (2023). \doiwoc{10.1103/PhysRevLett.130.051801}

\bibitem{losAlamos}
R.G.H. Robertson, T.J. Bowles, G.J. Stephenson, D.L. Wark, J.F. Wilkerson, D.A. Knapp, Limit on neutrino mass from observation of the beta decay of molecular tritium, Phys. Rev. Lett. \textbf{67}, 957 (1991). \doiwoc{10.1103/PhysRevLett.67.957}

\bibitem{mainz2005}
C.~Kraus, B.~Bornschein, L.~Bornschein, J.~Bonn, B.~Flatt, A.~Kovalik, B.~Ostrick, E.W. Otten, J.P. Schall, T.~Thümmler et~al., {Final results from phase II of the Mainz neutrino mass search in tritium ${\beta}$ decay}, The European Physical Journal C \textbf{40}, 447–468 (2005). \doiwoc{10.1140/epjc/s2005-02139-7}

\bibitem{troitsk2011}
V.N. Aseev, A.I. Belesev, A.I. Berlev, E.V. Geraskin, A.A. Golubev, N.A. Likhovid, V.M. Lobashev, A.A. Nozik, V.S. Pantuev, V.I. Parfenov et~al., {Upper limit on the electron antineutrino mass from the Troitsk experiment}, Physical Review D \textbf{84} (2011). \doiwoc{10.1103/physrevd.84.112003}

\bibitem{hardwarepaper}
M.~Aker, et~al., {The design, construction, and commissioning of the KATRIN experiment}, Journal of Instrumentation \textbf{16}, T08015 (2021). \doiwoc{10.1088/1748-0221/16/08/t08015}

\bibitem{katrin2022}
M.~Aker et~al. (KATRIN), {Direct neutrino-mass measurement with sub-electronvolt sensitivity}, Nature Phys. \textbf{18}, 160 (2022), \texttt{2105.08533}. \doiwoc{10.1038/s41567-021-01463-1}

\bibitem{katrin2024}
M.~Aker, D.~Batzler, A.~Beglarian, J.~Behrens, J.~Beisenkötter, M.~Biassoni, B.~Bieringer, Y.~Biondi, F.~Block, S.~Bobien et~al., {Direct neutrino-mass measurement based on 259 days of KATRIN data} (2024), \texttt{2406.13516}, \urlstyle{tt}\url{https://arxiv.org/abs/2406.13516}

\bibitem{oldp8}
A.~Ashtari~Esfahani et~al. (Project 8), {Determining the neutrino mass with cyclotron radiation emission spectroscopy\textemdash{}Project 8}, J. Phys. G \textbf{44}, 054004 (2017), \texttt{1703.02037}. \doiwoc{10.1088/1361-6471/aa5b4f}

\bibitem{p8PRL}
A.~Ashtari~Esfahani, S.~B\"oser, N.~Buzinsky, M.C. Carmona-Benitez, C.~Claessens, L.~de~Viveiros, P.J. Doe, M.~Fertl, J.A. Formaggio, J.K. Gaison et~al., {Tritium Beta Spectrum Measurement and Neutrino Mass Limit from Cyclotron Radiation Emission Spectroscopy}, Phys. Rev. Lett. \textbf{131}, 102502 (2023). \doiwoc{10.1103/PhysRevLett.131.102502}

\bibitem{ptolemy}
Y.~Iwasaki, A.~Tan, C.G. Tully (PTOLEMY), {Towards CRES-Based Non-destructive Electron Momentum Estimation for the PTOLEMY Relic Neutrino Detector} (2024), \texttt{2404.00817}

\bibitem{echo2019}
C.~Velte et~al., {High-resolution and low-background $^{163}$Ho spectrum: interpretation of the resonance tails}, Eur. Phys. J. C \textbf{79}, 1026 (2019). \doiwoc{10.1140/epjc/s10052-019-7513-x}

\bibitem{BORGHESI2023}
M.~Borghesi, B.~Alpert, M.~Balata, D.~Becker, D.~Bennet, E.~Celasco, N.~Cerboni, M.~{De Gerone}, R.~Dressler, M.~Faverzani et~al., An updated overview of the holmes status, Nuclear Instruments and Methods in Physics Research Section A: Accelerators, Spectrometers, Detectors and Associated Equipment \textbf{1051}, 168205 (2023). \doiwoc{https://doi.org/10.1016/j.nima.2023.168205}

\end{thebibliography}
\end{document}